\documentclass[pre,aps,floatfix,showpacs,reprint,onecolumn,superscriptaddress]{revtex4-1}

\usepackage{amssymb,amsfonts,amsmath}
\usepackage{graphics}
\usepackage[dvips]{graphicx}
\usepackage{color}
\usepackage{subfigure}
\usepackage{mathtools}
\usepackage[normalem]{ulem}
\usepackage{dcolumn}
\usepackage{multirow}
\usepackage{booktabs}
\usepackage{bm}
\usepackage{cancel}
\usepackage[linktocpage,colorlinks=true,linkcolor=blue,citecolor=blue]{hyperref}

\newcommand{\mf}[1]{\ensuremath{\mathbf{#1}}}

\newcommand{\mrm}[1]{\ensuremath{\mathrm{#1}}}

\begin{document}

\title{Hydrodynamic instabilities provide a generic route to \\ spontaneous biomimetic oscillations in chemomechanically active filaments}

\author{Abhrajit Laskar}
\affiliation{The Institute of Mathematical Sciences, CIT Campus, Chennai 600113, India}
\author{Rajeev Singh}
\affiliation{The Institute of Mathematical Sciences, CIT Campus, Chennai 600113, India}
\author{Somdeb Ghose}
\affiliation{The Institute of Mathematical Sciences, CIT Campus, Chennai 600113, India}
\author{Gayathri Jayaraman}
\affiliation{The Institute of Mathematical Sciences, CIT Campus, Chennai 600113, India}
\author{P. B. Sunil Kumar}
\affiliation{Department of Physics, Indian Institute of Technology Madras, Chennai 600036, India}
\author{R. Adhikari$^{\ast}$}
\affiliation{The Institute of Mathematical Sciences, CIT Campus, Chennai 600113, India \\
*Correspondence to rjoy@imsc.res.in}

\date{\today}

\maketitle

\section*{Abstract}

\textbf{
Non-equilibrium processes which convert chemical energy into mechanical motion enable the motility of organisms. Bundles of inextensible filaments driven by energy transduction of molecular motors form essential components of micron-scale motility engines like cilia and flagella. The mimicry of cilia-like motion in recent experiments on synthetic active filaments supports the idea that generic physical mechanisms may be sufficient to generate such motion. Here we show, theoretically, that the competition between the destabilising effect of hydrodynamic interactions induced by force-free and torque-free chemomechanically active flows, and the stabilising effect of nonlinear elasticity, provides a generic route to spontaneous oscillations in active filaments. These oscillations, reminiscent of prokaryotic and eukaryotic flagellar motion, are obtained without having to invoke structural complexity or biochemical regulation. This minimality implies that biomimetic oscillations, previously observed only in complex bundles of active filaments, can be replicated in simple chains of generic chemomechanically active beads.
}

\section*{Introduction}
Prokaryotic bacteria \cite{jahn1965} as well as eukaryotic sperm cells \cite{gray1955, lindemann1972} employ rhythmic flagellar beating for locomotion in viscous fluids. Bacterial flagella rotate rigidly in corkscrew fashion \cite{berg1973, berg2003}, while spermatic flagella behave more like flexible oars \cite{purcell1977} with their beating mostly confined to a plane \cite{brokaw1965, brennen1977, brokaw1991}. 
Oscillatory motility in clamped flagella can arise spontaneously and, with an unlimited supply of energy, can persist indefinitely without any external or internal regulatory pacemaker mechanism \cite{lindemann1972, fujimura2006}. Autonomous motility as well as spontaneous beating due to hydrodynamic instabilities has been recently reproduced \emph{in vitro} \cite{sanchez2011, sanchez2012}, where a biomimetic active motor-microtubule assemblage has been shown to exhibit remarkable cilialike beating motion with hydrodynamic interactions (HI) playing a crucial role in synchronised oscillations \cite{sanchez2011}. Previous models \cite{machin1958, brokaw1971, lighthill1976, hines1978, gueron1992, lindemann1994, camalet1999, camalet2000, dillon2000, riedel-kruse2007, kikuchi2009, spagnolie2010} analysing the mechanism behind flagellar beating have, in general, ignored the role of HI.

\begin{figure*}[tbp]  
 \begin{center}
  \includegraphics[width=0.99\textwidth]{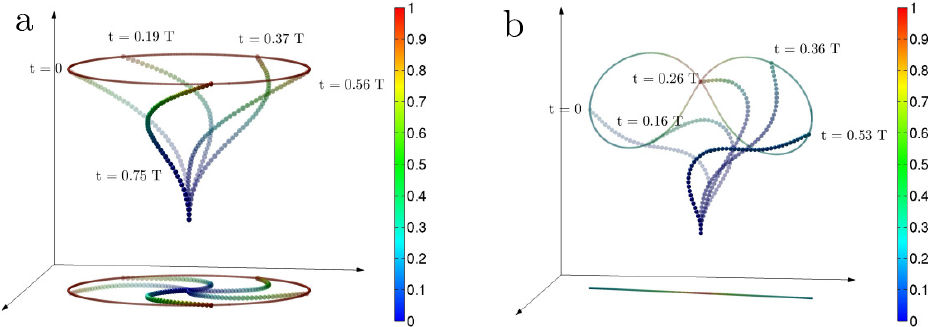}
  \caption{Biomimetic oscillations of the clamped filament plotted at different times over an oscillation period $T$. 
  In (a) we see rigid aplanar corkscrew rotation for $\mathcal{A} = 25$ while in (b) we see flexible planar beating for $\mathcal{A} = 50$.
  The colour of the beads as well as the trace of the tip correspond to individual instantaneous monomer speeds. The colourbars are normalised by the maximum speed.}
 \end{center}
\end{figure*}

\begin{figure*}[tbp]  
 \begin{center}
  \includegraphics[width=0.95\textwidth]{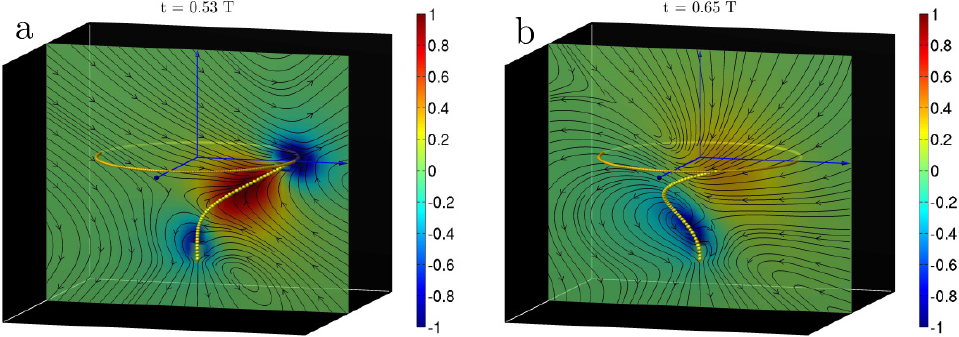}
  \caption{Flow fields of rigid aplanar corkscrew rotation at two different instances of rotation. The colour indicates the signed magnitude of the velocity field perpendicular to the plane normalised by its maximum.}
 \end{center}
\end{figure*}

\begin{figure*}[h!]  
 \begin{center}
  \includegraphics[width=0.95\textwidth]{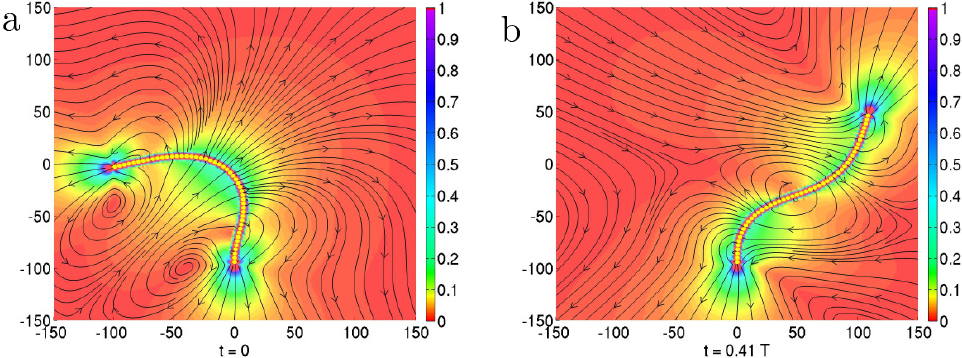}
  \caption{Flow fields of flexible planar beating at two instants of the oscillation cycle. The colour indicates the magnitude of the velocity in the plane normalised by its maximum.}
 \end{center}
\end{figure*}

\begin{figure*}[tbp]  
 \begin{center}
  \includegraphics[width=0.98\textwidth]{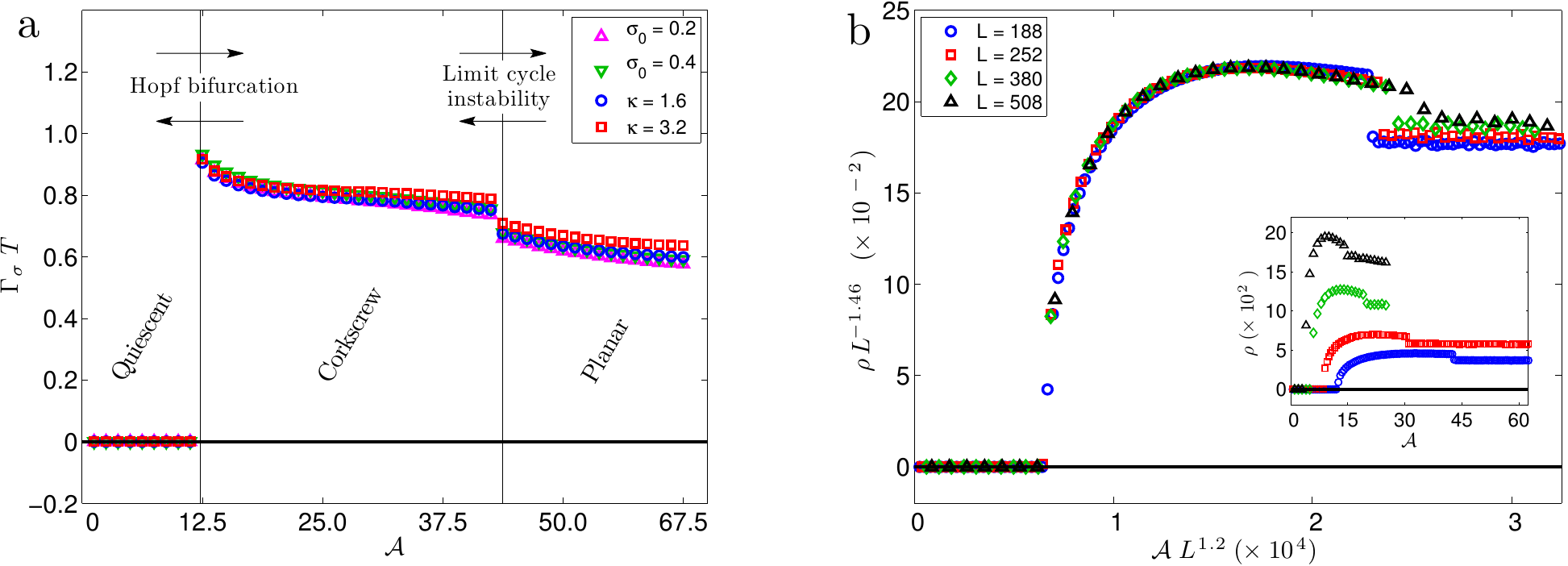}
  \caption{Variation of the scaled timeperiod $\Gamma_{\sigma} T$ of filament beating with $\mathcal{A}$ plotted for various values of $\kappa$ and $\sigma_0$ with $L = 188$, and (b, main panel) variation of the scaled amplitude $L^{-1.46} \rho $ with $\mathcal{A}$ plotted for various lengths $L$ with $\kappa = 1.6$.
In (a) we show the appearance of spontaneous oscillations in the filament at $\mathcal{A}\sim 12.5$ corresponding to rigid corkscrew rotation, followed by a transition at $\mathcal{A}\sim 45$ to flexible planar beating. In (b, inset) we show the increase in the unscaled amplitude with length.}
 \end{center}
\end{figure*}

\begin{figure*}[htp]  
 \begin{center}
  \includegraphics[width=0.98\textwidth]{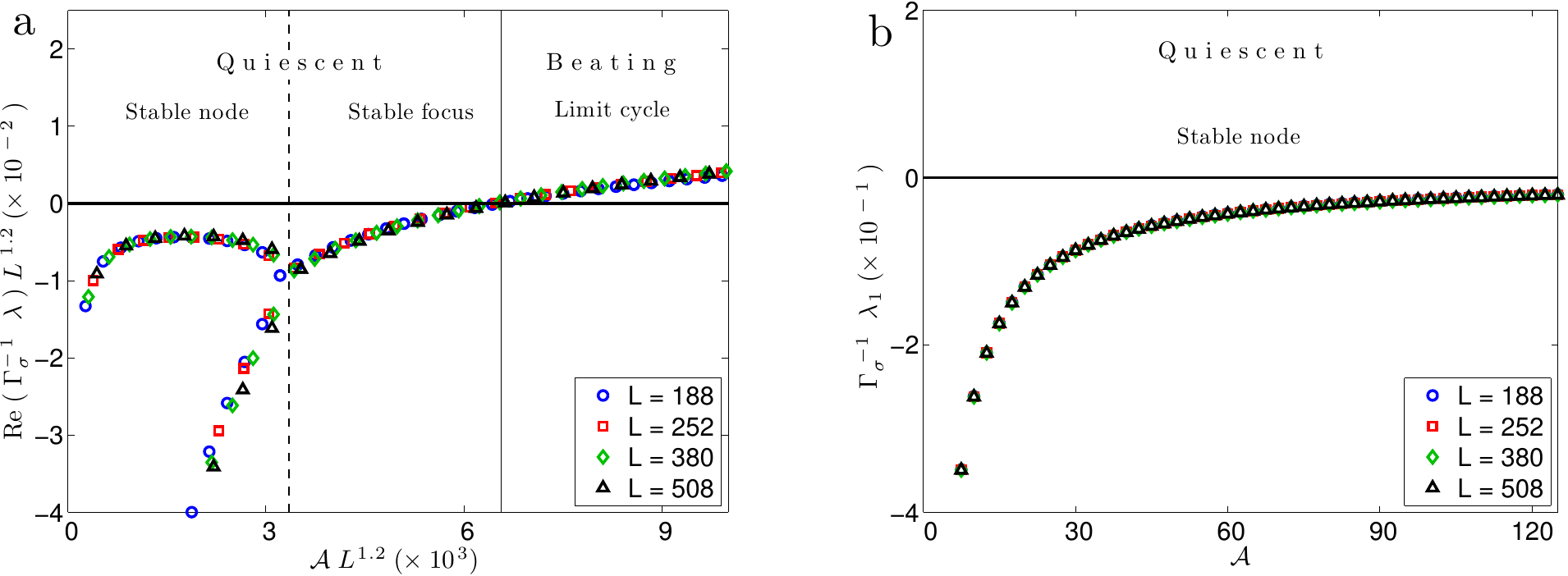}
  \caption{Variation of the largest scaled eigenvalues of the Jacobian matrix (a) including HI and (b) excluding HI. In (a) we see the transition from a stable node to a stable focus followed by a supercritical Hopf bifurcation from quiescence to a limit cycle. In (b) we see only a stable quiescent state. In the absence of HI the eigenvalue scaling is completely determined by $\mathcal{A}$. Comparing (a) and (b) it is clear that hydrodynamic instabilities due to HI are the main mechanism for spontaneous oscillations. (a) also shows that the rate of relaxation (given by the magnitude of the real part of largest eigenvalue) to the steady state before the bifurcation decreases as we approach the bifurcation from below.}
 \end{center}
\end{figure*}
%
Here we study a minimal active filament model \cite{jayaraman2012} which, once clamped at one end, exhibits a variety of spontaneous beating phenomena in a three dimensional fluid. Our model filament consists of chemomechanically active beads (CABs) which convert chemical energy to mechanical work in viscous fluids. These CABs are connected through potentials that restricts extensibility and enforces semiflexibility and self-avoidance of the filament. The conversion of chemical energy to mechanical work within the fluid produces flows which do not add net linear or angular momentum to it and, thus, must be represented at low Reynolds numbers by force-free and torque-free singularities \cite{blake1971a, brennen1977, lauga2009, ramaswamy2010, cates2011, marchetti2012}. We model the activity of the beads by a stresslet singularity which produce a flow decaying as $1/r^2$. This stresslet contribution arises from chemomechanical activity, for instance the metachronal waves of ciliated organisms \cite{sanchez2011}, or from phoretic flows in synthetic catalytic nanorods \cite{paxton2004, vicario2005, ozin2005, catchmark2005}. 
For self-propelled particles, additional dipolar contributions generating flows decaying as $1/r^3$ are present, but are neglected here as they are subdominant to stresslet contributions. The equation of motion for the active filament \cite{jayaraman2012} incorporating the effects of nonlinear elastic deformations, active processes and HI is
%
%
\begin{align}\label{eq:activeEoM}
\dot {\bf r}_n = \sum_{m=1}^N\left [ {\mathbf O}({\bf r}_n - {\bf r}_m)\cdot {\bf f}_m+ {\mathbf D}({\bf r}_n - {\bf r}_m)\cdot{\boldsymbol \sigma}_m \right]
\end{align}
%
%
where ${\bf r}_n$ is the location of the $n$-th bead, ${\bf f}_n$ is the total elastic force on the $n$-th bead, and ${\boldsymbol \sigma}_n = \sigma_0({\bf t}_n {\bf t}_n - \mathbb{I}/3)$ is stresslet tensor directed along the the local unit tangent ${\bf t}_n$. Here $\sigma_0 > 0$ sets the scale of (extensile) activity. The monopolar Oseen tensor $\mathbf{O}$ and the dipolar stresslet tensor $\mathbf{D}$ respectively propagate the elastic and active contributions to the flow (details of model in Supplementary Text). Noise, of both chemomechanical and thermal origin can be added to these equations, but are not considered here. We impose clamped boundary conditions at one end and solve the equation of motion through direct summation of the hydrodynamic Green's functions. For a filament of length $L$ and bending modulus $\kappa$ the dynamics is characterised by the dimensionless activity number $\mathcal{A} = {L\sigma_0 / \kappa}$ \cite{jayaraman2012}.

\section*{Results}

\textbf{Spontaneous oscillations.} 
We briefly recall the mechanism behind hydrodynamic instabilities in active filaments  \cite{jayaraman2012}. Extensile activity in a straight filament produces flows with dipolar symmetry that point tangentially outward at the filament ends and normally inward at the filament midpoint. A spontaneous transverse perturbation breaks flow symmetry about the filament midpoint resulting in a net flow in the direction of the perturbation. The destabilising effect of the hydrodynamic flow is countered by the stabilising effect of linear elasticity for activity numbers $\mathcal{A} < \mathcal{A}_{c1}$ but leads to a linear instability for $\mathcal{A} > \mathcal{A}_{c1}$. This instability produces filament deformations which are ultimately contained by the non-linear elasticity producing autonomously motile conformations \cite{jayaraman2012}. Here, the additional constraint imposed by the clamp transforms the autonomously motile states into ones with spontaneous oscillations. We perform numerical simulations of the active filament model to show that the interplay of hydrodynamic instabilities, non-linear elasticity, and the constraint imposed by the clamp leads to spontaneously oscillating states.

Numerical simulations of Eq.\ (\ref{eq:activeEoM}) reveal two distinct oscillatory states (Figs.\ 1a, 1b, Supplementary Fig.\ S3a, Supplementary Videos 1 and 2). The first of these, seen in the range $\mathcal{A}_{c1} < \mathcal{A} < \mathcal{A}_{c2}$, is a state in which the filament rotates rigidly in a corkscrew-like motion about the axis of the clamp. This rotational corkscrew motion is reminiscent of prokaryotic flagellar beating \cite{berg1973, berg2003}. We show this motion in Fig.\ 1a over one time period of oscillation together with the projection of the filament on the plane perpendicular to the clamp axis. A section of the three-dimensional flow in a plane containing the clamp axis is shown in Figs.\ 2a and 2b. The net flow points in the direction opposite to the filament curvature and the entire flow pattern co-rotates with the filament. In the second state, seen for $\mathcal{A} > \mathcal{A}_{c2}$, the filament beats periodically in a two-dimensional plane containing the axis of the clamp, with waves propagating from the clamp to the tip. This flexible beating  is reminiscent of eukaryotic flagellar motion \cite{gray1955, brokaw1965, lindemann1972, brokaw1991, fujimura2006}. We show this motion in Fig.\ 1b over one time period of oscillation together with the projection of the filament on the plane perpendicular to the clamp axis. The projection is now a line, showing that motion is confined to a plane. A section of the three-dimensional flow in the plane of beating is shown in Figs.\ 3a and 3b. Two distinct types of filament conformations of opposite symmetry are now observed, corresponding to different parity of the conformation with respect to the perpendicular bisector of the line joining the two end points. In the  \emph{even} conformation (Fig.\ 3a), the flow points in the direction opposite to the curvature as in the corkscrew state. However, in the \emph{odd} conformation (Fig.\ 3b), the flow has a centre of vorticity at the point of inflection of the filament. This centre of vorticity moves up the filament and is shed at the tip at the end of every half cycle. The critical activities scale as $\mathcal{A}_{c1} = L^{-1.2}$ and $\mathcal{A}_{c2} = L^{-1.1}$, obtained from a Bayesian parameter estimation of data shown in Fig.\ S3b. The critical values depend only on the ratio $\sigma_0 / \kappa$, and not on $\sigma_0$ and $\kappa$ individually, as is clearly seen in Fig.\ 3a.

\textbf{Time period and amplitude scaling.} 
The physical parameters determining the time period $T$ of the oscillatory states are the active stresslet $\sigma_0$, the bending modulus $\kappa$, the fluid viscosity $\eta$ and the filament length $L$. Remarkably, variations of $T$ in this four-dimensional parameter space collapse, when scaled by the active relaxation rate $\Gamma_{\sigma} = \sigma_0/\eta L^3$, to a one-dimensional scaling curve of the form $L^{-\alpha}f(\mathcal{A}/L^{\beta})$. We show the data collapse at fixed $L$ and varying relative activity $\sigma_0/\kappa$ in Fig.\ 4a while the scaling with system size is shown in Supplementary Fig.\ S2a. Our best estimates for the exponents, obtained from Bayesian regression, are $\alpha = 1.3$ and $\beta = -1.2$ (Supplementary Fig.\ S2a). Qualitatively, at fixed relative activity the oscillation frequency decreases with increasing $L$, while at a constant $L$ the oscillation frequency increases with increasing relative activity. 
This is in agreement with a simple dimensional estimate of the time period $T \sim \eta L^3 / \sigma_0$. For active beads with a stresslet of $\sigma_0 \sim  6 \times 10^{-18} Nm$ in a filament of length $L\sim 100\mu m$, our estimate of the time period gives a value of $170 s$, which agrees in order of magnitude with experiment \cite{sanchez2011}. The amplitude of oscillation $\rho$ obeys a similar scaling relation with $\alpha = - 1.46, \beta = - 1.2$ (Fig.\ 4b, main panel). At fixed relative activity, $\rho$ increases with increasing $L$, while at fixed $L$, it increases and then saturates at large relative activity. The amplitude in the planar beating state is marginally smaller than in the corkscrew rotating state (Fig.\ 4b, inset).

\textbf{Linear Stability and Hopf Bifurcation.} 
To better understand the nature of the hydrodynamic instability and the transition to spontaneous oscillation we performed a linear stability analysis \cite{strogatz1994} of the straight filament. In absence of activity, $\mathcal{A} = 0$, all eigenvalues of the Jacobian are real and negative and the filament has an overdamped relaxation to equilibrium. With increasing $\mathcal{A}$, the two largest real eigenvalue pairs approach, converge, and become complex conjugate pairs. This corresponds to a transition from a stable node to a stable focus where the response changes from being overdamped to underdamped. The analysis reveals that the balance between hydrodynamic flow and linear elasticity has a non-monotonic variation. While the general trend is towards slower relaxations with increasing $\mathcal{A}$ corresponding to the greater relative strength of the hydrodynamic flow, this is reversed in a small window of activity where the increasing activity produces faster relaxations. This can be clearly seen in Fig.\ 5a and Fig.\ S1a, where the rate of relaxation is given by the magnitude of the real part of the largest eigenvalue. With further increase of $\mathcal{A}$ the complex eigenvalues approach the imaginary axis monotonically, crossing them at a critical value $\mathcal{A}_{c1}$ (Fig.\ 5a, Supplementary Figs.\ S1a and S1b, and Supplementary Video 3). Through this supercritical Hopf bifurcation, the stable focus flows into the limit cycle corresponding to the corkscrew rotation. The value of $\mathcal{A}_{c1}$ obtained from the linear stability analysis is in perfect agreement with that obtained from numerical simulation. As with the time-period and amplitude, the eigenvalues $\lambda$ of the Jacobian obey scaling relations $\lambda \Gamma_{\sigma}^{-1} = L^{-\alpha} f(\mathcal{A}/L^{\beta} )$ with $\alpha = 1.2$ and $\beta = -1.2$.

\textbf{Importance of HI.}
To ascertain the importance of HI, we repeat the stability analysis  on a local limit of our model. Here, the long-ranged contributions to the hydrodynamic flow from both elasticity and activity are neglected and only their short-ranged effects are retained (see Supplementary Text). We find that all eigenvalues remain real and negative for activity numbers corresponding to an order of magnitude greater than $\mathcal{A}_{c1}$, reflecting the stability of the quiescent state in the absence of HI (Fig.\ 5b). 

\section*{Discussions}

Our work shows that simple chains of CABs, for instance of synthetic catalytic nanorods \cite{paxton2004, vicario2005, ozin2005, catchmark2005}, can show the spontaneous beating obtained previously in more complex systems like self-assembled motor-microtubule mixtures \cite{sanchez2011} or externally actuated artificial cilia \cite{dreyfus2005, evans2007, vilfan2010, coq2011}. We emphasise that an experimental realisation of our system requires neither external actuation nor self-propulsion. The only chemomechanical requirement is that the CABs produce force-free and torque-free dipolar flows in the fluid. This makes them an attractive candidate for biomedical applications like targetted drug delivery. Our detailed prediction for the spatiotemporal dynamics of the hydrodynamic flows can be experimentally verified using particle imaging velocimetry \cite{drescher2010}.

In summary, we have shown that a minimal filament model which includes elasticity, chemomechanical activity and HI, exhibits spontaneous emergent biomimetic behaviour reminiscent of the rhythmic oscillations of various prokaryotic and eukaryotic flagella \cite{jahn1965, gray1955, lindemann1972, berg1973, berg2003, purcell1977, brokaw1965, brokaw1991}. Our results lead us to conclude that hydrodynamic instabilities due to internal active stresses are sufficient to induce spontaneous biomimetic beating in a clamped chemomechanically active filament.

\section*{Methods}

We calculate the RHS of Eq.\ (\ref{eq:activeEoM}) by a direct summation of the hydrodynamic Green's functions. Clamped boundary conditions are implemented at one end by fixing the position of the first particle and allowing the second particle to move only along the tangential direction.
The equation of motion is integrated using a variable step method as implemented in ODE15s in Matlab. The hydrodynamic flows fields are obtained on a regularly spaced Eulerian grid by summing the individual contributions from each of the $N$ particles. The linear stability analysis is performed by first numerically integrating the equations of motion to obtain the fixed point, then numerically evaluating the $3N\times 3N$ Jacobian matrix at the fixed point, and finally computing the eigenvalues of the  Jacobian matrix numerically. Simulations are carried out for different filament lengths $L$ with bead numbers upto $N = 128$. The equilibrium bond length is taken to be $b_0 = 4$. We choose $\kappa$ in the range $0.0$ to $1.0$ and $\sigma_0$ in the range $0.0$ to $0.5$. The initial condition is a random transverse perturbation applied to every particle.
Random perturbations in the longitudinal direction relaxes at a much faster time-scale due to the stretching potential. The total integration time is typically $10\Gamma_{\sigma}^{\, \, -1}$, where $\Gamma_{\sigma} = \sigma_0 / \eta L^3$ is the active relaxation rate, and $\eta$ is the viscosity, taken to be $1/6$.

\section*{Acknowledgements}
Financial support from PRISM II, Department of Atomic Energy, Government of India and computing resources through HPCE, IIT Madras and Annapurna, IMSc are gratefully acknowledged. We thank M. E. Cates, Z. Dogic, D. Frenkel, G. Baskaran, I. Pagonabarraga, and R. Simon for helpful discussions, and P. V. Sriluckshmy for help with Bayesian analysis.

\section*{Author Contributions}
R.A. and P.B.S.K. designed research. A.L., R.S., S.G. and G.J. performed research. S.G., R.A., R.S. and P.B.S.K. wrote the manuscript.

\section*{Additional information}
\textbf{Competing financial interests.}
The authors declare no competing financial interests.

\section*{Supplementary Information}

\subsection*{Supplementary : Model}

\noindent
Following Ref.\ \cite{jayaraman2012}, we construct an active elastic filament by chaining active beads using potentials. We place $N$ such beads at points $\mf{r}_1, \mf{r}_2, \ldots \mf{r}_N$ and define bond vectors ${\bf b}_m = {\bf r}_{m+1} - {\bf r}_m$ between adjoining beads. The potentials $U_{\rm S} ({\bf b}_{m}) = \frac{1}{2} k(|{\bf b}_{m}| - b_{0})^2$  and $U_{\rm B}({\bf b}_{m}, {\bf b}_{m+1}) = (\kappa/b_0)(1- \cos\phi_{m})$ model inextensibility and semiflexibility respectively, penalizing departures of the filament from the equilibrium bead-bead separation of $b_0$ or the bond-bond angle $\phi_m = 0$. Here $k$ is the spring constant and $\kappa$ is the bending modulus. Self-avoidance is enforced through a purely repulsive Lennard-Jones potential $U_{LJ}$ which vanishes smoothly at a distance $\sigma_{LJ}$. The total potential $U({\bf r}_1, \ldots, {\bf r}_N)$ is the sum of these three potentials. Stretching, bending and self-avoidance causes the total elastic force ${\bf f}_n = -\partial U({\bf r}_1, \ldots, {\bf r}_N) / \partial {\bf r}_n$ to act on the $n$-th bead. 


We model the activity using force-free and torque-free singularities \cite{blake1971a, brennen1977, ramaswamy2010, cates2011, marchetti2012}, of which the second-rank symmetric stresslet tensor is the most dominant \cite{chwang1975}, and produces flows that decay inverse squarely with distance. With a tensorial strength $\bm{\sigma}_n$ and an axis of uniaxial symmetry $\mf{p}_n$, the stresslet can be parametrised as $\bm{\sigma}_n = \sigma_0({\bf p}_n{\bf p}_n - \mathbb{I}/d)$ where $d$ is the spatial dimension and $\sigma_0$ sets the scale of the activity. 
Extensile flows correspond to $\sigma_0 > 0$, while contractile flows correspond to $\sigma_0 < 0$. In the present case, we set $\mf{p}_n = \hat{\mf{t}}_n$, the local unit tangent vector, to reflect the tangential stresses generated by the active particles.


The filament exerts forces on the surrounding fluid due to its elasticity and activity. The resultant force density at the point $\mf{r}$ due to $N$ active beads is given by summing over the Stokeslet and stresslet singularities,
\begin{align}\label{eq:activeF}
{\bf F}({\bf r}) = 
\sum_{m} \left\{ {\bf f}_m \delta({\bf r} - {\bf r}_m)
+ \nabla\cdot\left[{\boldsymbol \sigma}_m \delta({\bf r} - {\bf r}_m)\right] \right\}
\end{align}
Integrating this on a surface enclosing the filament gives 
\begin{equation}
\int d^3{\bf r}\, {\bf F}({\bf r}) = 
\sum_m \left\{{\bf f}_m + \int d^3{\bf r}\, \nabla \cdot \left[{\boldsymbol \sigma}_m \delta({\bf r} - {\bf r}_m)\right] \right\} = 0.
\end{equation}
Since the elastic forces $\mf{f}_m$ are internal to the filament and obey Newton's third law, they  cancel exactly on summation over all the beads. The active terms are total divergences and thus vanish on integration over a bounding surface. The symmetry of the stresslet tensor ensures that no angular momentum is added to the fluid. Consequently, the force density considered here does not add any \emph{net} linear or angular momentum to the fluid. Models in which this fundamental constraint is not imposed will fail to correctly reproduce HI due to active energy transduction. 


In the low Reynolds number regime, a force $\mf{f}(\mf{r}')$ and a stresslet $\bm{\sigma}(\mf{r}')$, placed at location $\mf{r}'$ in a three-dimensional unbounded fluid, produce flows 
$v_{\alpha}^{\, \mrm{el}} (\mf{r}) = O_{\alpha\beta}(\mf{r} - \mf{r}') f_{\beta} (\mf{r}')$ 
and $v_{\alpha}^{\mrm{ ac}}(\mf{r}) = D_{\alpha\beta\gamma}(\mf{r} - \mf{r}') \sigma_{\beta\gamma} (\mf{r}')$ respectively. The Oseen and stresslet tensors are given by $O_{\alpha\beta}({\bf r}) = (\delta_{\alpha\beta} + \hat{r}_{\alpha} \hat{r}_{\beta})/8 \pi \eta r$ and $D_{\alpha\beta\gamma}({\bf r}) = (-\hat{r}_{\alpha} \delta_{\beta\gamma} + 3 \hat{r}_{\alpha} \hat{r}_{\beta} \hat{r}_{\gamma})/ 8 \pi \eta r^2$ \cite{pozrikidis1992, kim2005} where $\hat{r}_{\alpha} = {\bf r}_{\alpha}/r$  and $\alpha$, $\beta$ are the Cartesian coordinates.

The velocity of the $n$-th bead is obtained by summing the force and activity contributions from all beads, including itself, to the fluid velocity at its location. Thus, we obtain the equation of motion \cite{jayaraman2012}
\begin{align}
\dot {\bf r}_n = \sum_{m=1}^N\left [ {\mathbf O}({\bf r}_n - {\bf r}_m)\cdot {\bf f}_m+ {\mathbf D}({\bf r}_n - {\bf r}_m)\cdot{\boldsymbol \sigma}_m \right].
\end{align}
An isolated spherical bead with a force ${\bf f}$ acquires a velocity $\mu \, {\bf f}$ where $\mu$ is its mobility. By symmetry, an isolated spherical bead with a stresslet $\boldsymbol \sigma$ cannot acquire a velocity. Therefore, for $m=n$, $ O_{\alpha\beta} = \mu\delta_{\alpha\beta}$ and $D_{\alpha\beta\gamma} = 0$.
In the absence of activity, ${\boldsymbol \sigma}_n = 0$, and bending, $\kappa = 0$, the equation reduces to the Zimm model of hydrodynamic interactions of a polymer in a good solvent 
\cite{doi1988,*zimm1956}. 
Dimensionally, the active and elastic forces are of the form $\sigma_0/L$ and $\kappa/L^2$, where $L = (N-1)b_0$ is the length of the filament. The balance of these forces gives the dimensionless quantity $\mathcal{A} = L \sigma_0 / \kappa$, which is also the ratio of the active and elastic rates of relaxation, respectively $\Gamma_{\sigma} = \sigma_0/\eta L^d$ and $\Gamma_{\kappa} = \kappa/\eta L^{d+1}$ \cite{jayaraman2012}. The dynamics of the filament is completely captured by its length $L$ and the relative active strength, given by this activity number $\mathcal{A}$.

In the free-draining approximation to our model, we ignore HI. Thus the velocity of the n-th bead due to elastic forces is $\mu \mf{f}_n$, where $\mu$ is the mobility. For the active velocity we retain contributions from immediate neighbours of the n-th bead. This gives a local equation of motion 
\begin{align}
  \dot{\mf{r}}_n 
  &= \mu \mf{f}_n + \frac{\sigma_0}{4 \pi \eta b_0^3} \left( \mf{b}_{n} - \mf{b}_{n-1} \right)
\end{align}
The parameter values used for the simulation and analysis are: fluid viscosity $\eta = 1/6$, radius of monomer $a = 1$,  bond length $b_0 = 4a = 4$, spring constant $k = 1$ and the Lennard-Jones parameters $e=0.001, r_{min}=b=1$. Number of monomers varied from $N=24$ to $N=128$, while the remaining parameters $\kappa$ is chosen in the range $0.0$ to $1.0$ and $\sigma_0$ in the range $0.0$ to $0.5$.

\subsection*{Supplementary : Video titles and captions}

\noindent
\textbf{Video S1} : \emph{Aplanar corkscrew-like rigid rotation of clamped active filament with flowfield}.

\smallskip
\noindent
Description : Filament motion for $L = 188$ and $\mathcal{A} = 25$, displaying rigid corkscrew-like rotation about the axis of the clamp over three time periods of oscillation. The time trace of the filament tip as well as a section of the three-dimensional flow in a plane containing the clamp axis are shown. The net flow points in the direction opposite to the filament curvature and the entire flow pattern co-rotates with the filament. This motion is reminiscent of prokaryotic flagellar beating.

\bigskip
\noindent
\textbf{Video S2} : \emph{Planar flexible periodic beating of clamped active filament with flowfield}.

\smallskip
\noindent
Description : Filament motion for $L = 188$ and $\mathcal{A} = 50$, displaying flexible periodic beating in a two-dimensional plane containing the axis of the clamp over two time periods of oscillation. A section of the three-dimensional flow in the plane of beating is shown. Two distinct types of filament conformations of opposite symmetry are observed as the filament oscillates. In the \emph{even} conformation  the flow points in the direction opposite to the curvature as in the corkscrew state. However, in the \emph{odd} conformation the flow has a centre of vorticity at the point of inflection of the filament. This centre of vorticity moves up the filament and is shed at the tip at the end of every half cycle.

\bigskip
\noindent
\textbf{Video S3} : \emph{Hopf bifurcation in clamped active filament}.

\smallskip
\noindent
Description : Variation of real and imaginary parts of eigenvalues with activity number $\mathcal{A}$ for $L = 188$. The main panel shows the two largest eigenvalue pairs (red and blue pentagons) while the inset shows the entire spectrum. All eigenvalues are real and negative for $\mathcal{A} < 6$, beyond which the first pair converge and become complex conjugates, indicating the transition from the stable node to the stable focus. This pair crosses the imaginary axis at $\mathcal{A} \approx 12.5$, that is, at $\mathcal{A}_{c1}$, indicating the transition from the stable focus to a limit cycle through a Hopf bifurcation. The second pair replicates this entire behaviour at higher $\mathcal{A}$.

\subsection*{Supplementary : Figures}
\begin{figure*}[Hbp]    
 \begin{center}
 \subfigure[~]{\includegraphics[width=0.49\textwidth]{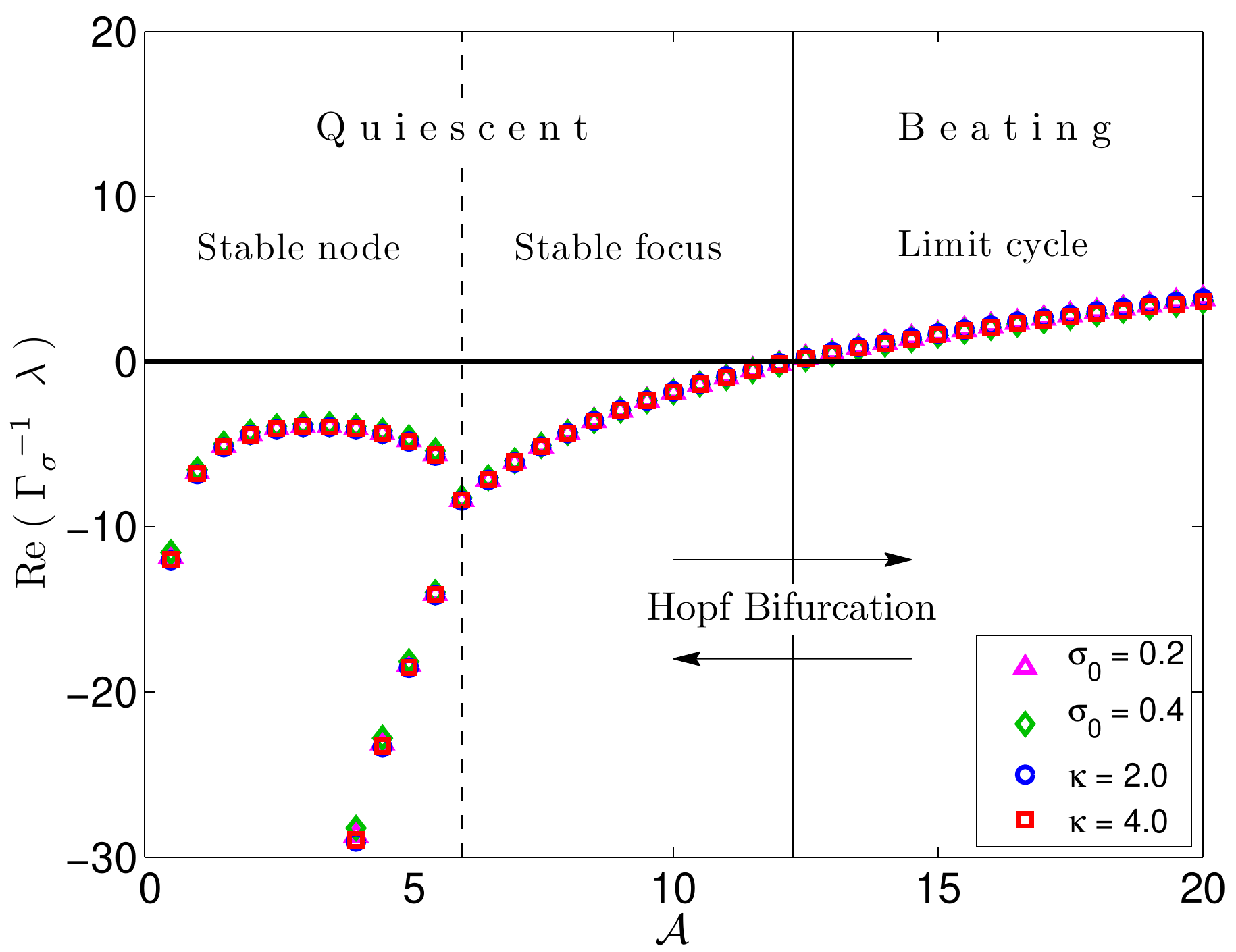} \label{fig:eigenvalue_scaling_N_48}} 
 \subfigure[~]{\includegraphics[width=0.49\textwidth]{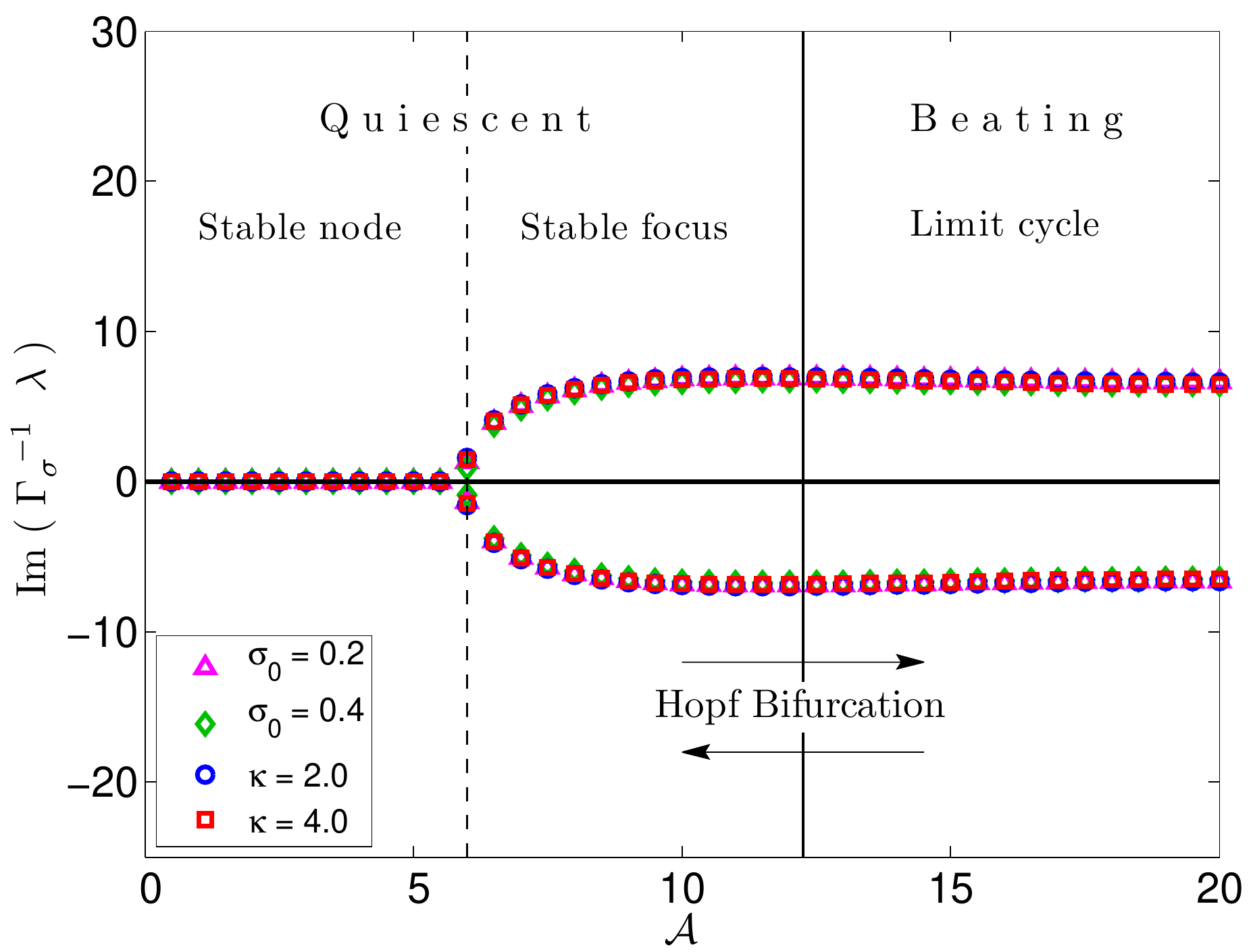} \label{fig:eigenvalue_imag_scaling_N_48}}
 \caption{FIG.\ S1. Variation of the scaled (a) real and (b) imaginary parts of the largest eigenvalues with $\mathcal{A}$, plotted for various values of $\kappa$ and $\sigma_0$ with $L = 188$. Data is obtained from a linear stability analysis (LSA) of the filament model. The largest eigenvalue pairs converge at $\mathcal{A} \sim 6$ and become complex with negative real parts, signalling the transition from stable node to stable focus. Re($\lambda$) become positive at $\mathcal{A}_{c1} \sim 12.5$ while Im($\lambda$) varies smoothly with $\mathcal{A}$, indicating a Hopf bifurcation into a limit cycle.
 } 
 \end{center}
\end{figure*}

\begin{figure*}[tbp]    
 \begin{center}
  \subfigure[~]{\includegraphics[width=0.49\textwidth]{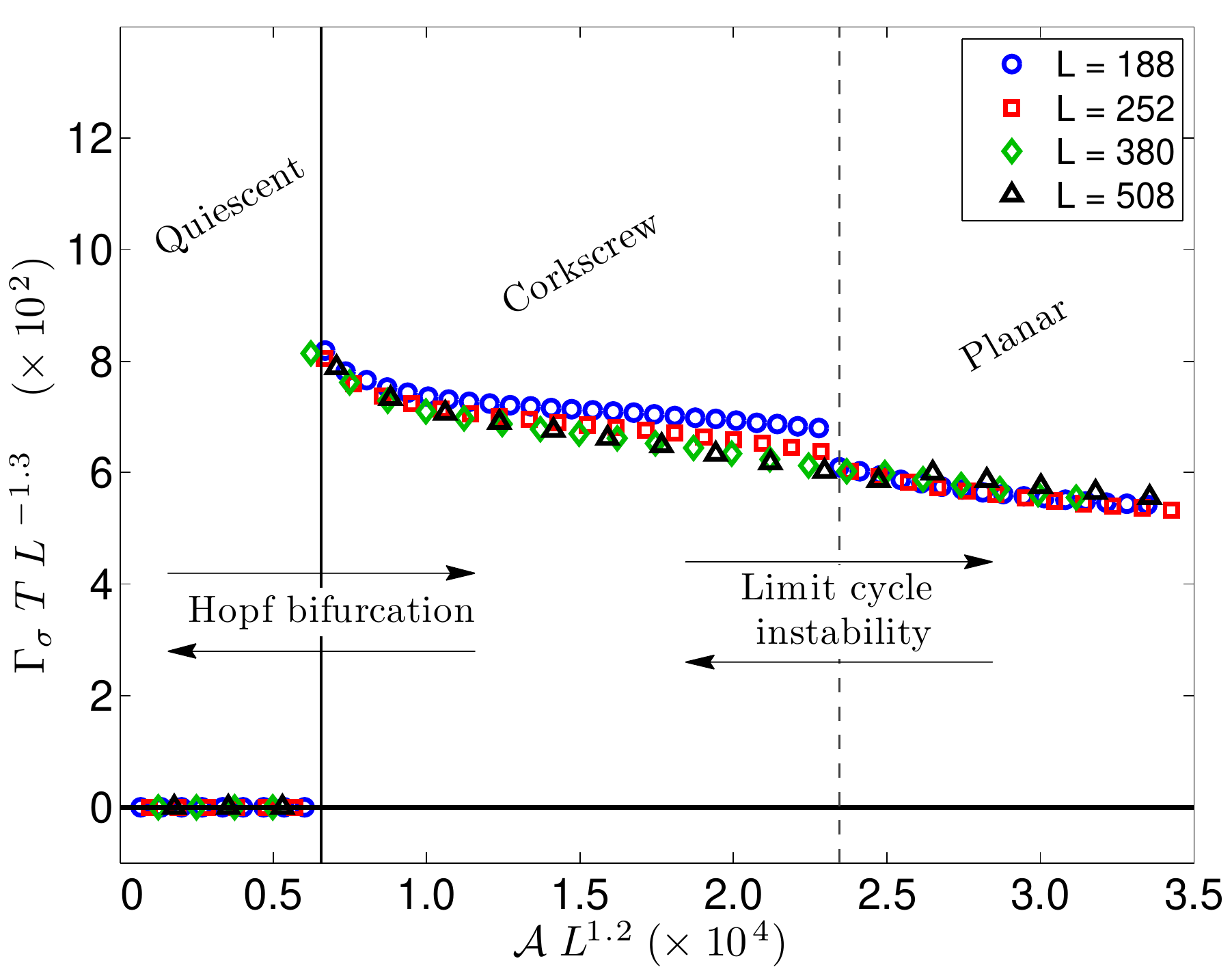} \label{fig:timeperiod_L_scaling_annotated}}
 \subfigure[~]{\includegraphics[width=0.49\textwidth]{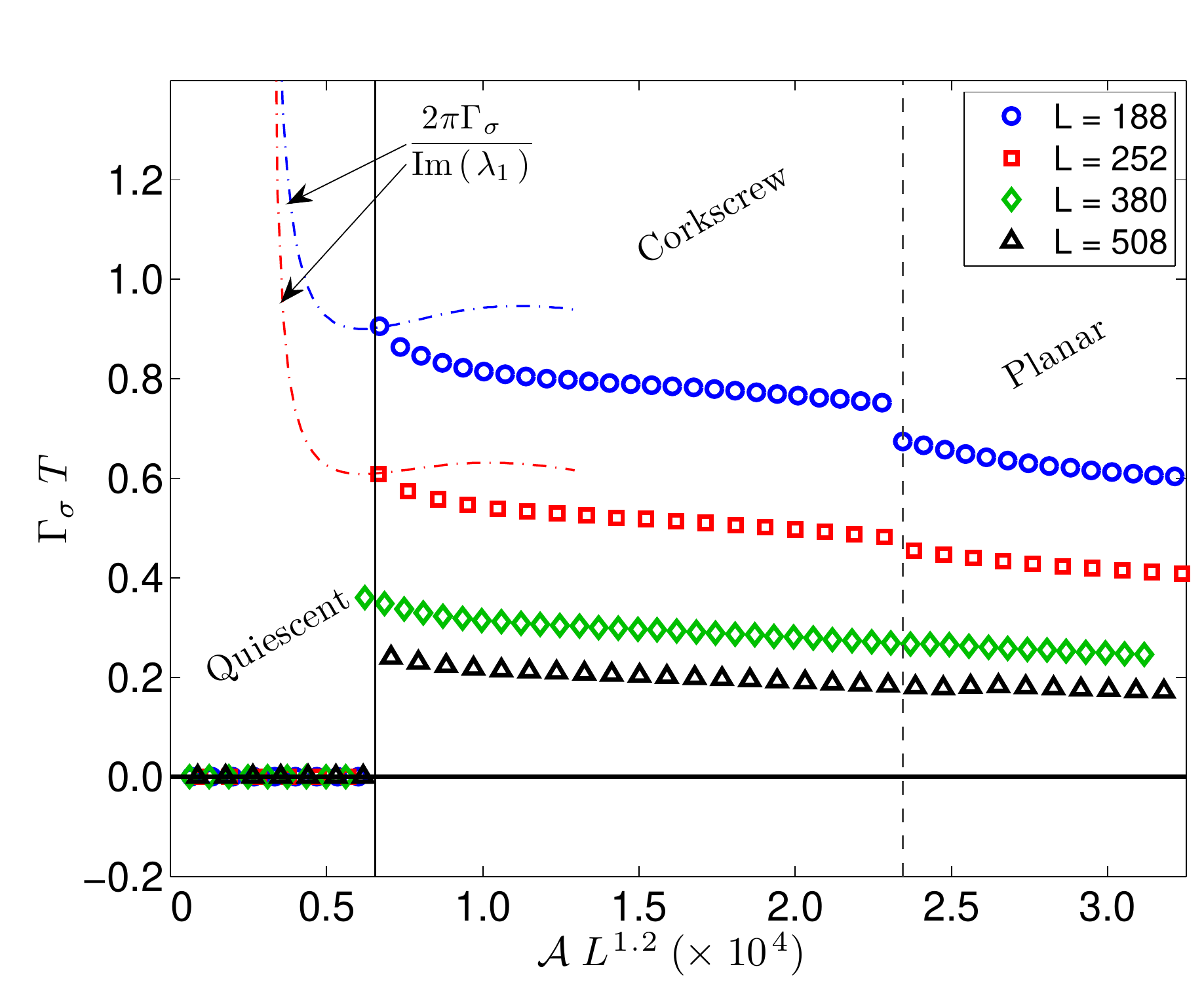} \label{fig:timeperiod_GammaScaling_L_variation}} 
 \caption{FIG.\ S2. Variation of the nondimensionalised time period $\Gamma_{\sigma} T$ with activity number $\mathcal{A}$ (a) with and (b) without $L$ scaling, obtained from numerical simulations. The dashed  lines in (b) represents the variation of $2 \pi \Gamma_{\sigma} / \mrm{Im} (\lambda_1)$ with $\mathcal{A}$ for $L = 188$ and $L = 252$, data obtained from LSA. 
 The rescaled plot of $\Gamma_{\sigma} T$ in (a) shows that its variations are well captured by a scaling form $L^{-\alpha}f(\mathcal{A} / L^{\beta})$, with $\alpha = 1.3$ and $\beta = -1.2$ estimated using Bayesian regression.
 The unscaled results in (b) shows that the time period is of the order of the active timescale $\Gamma_{\sigma}^{-1}$. The LSA estimate of the time period and the simulation result agree very well near the Hopf bifurcation point, and, predictably, deviates in the nonlinear regime. 
 } 
 \end{center}
\end{figure*}

\begin{figure*}[tbp]    
 \begin{center}
  \subfigure{\includegraphics[width=0.47\textwidth]{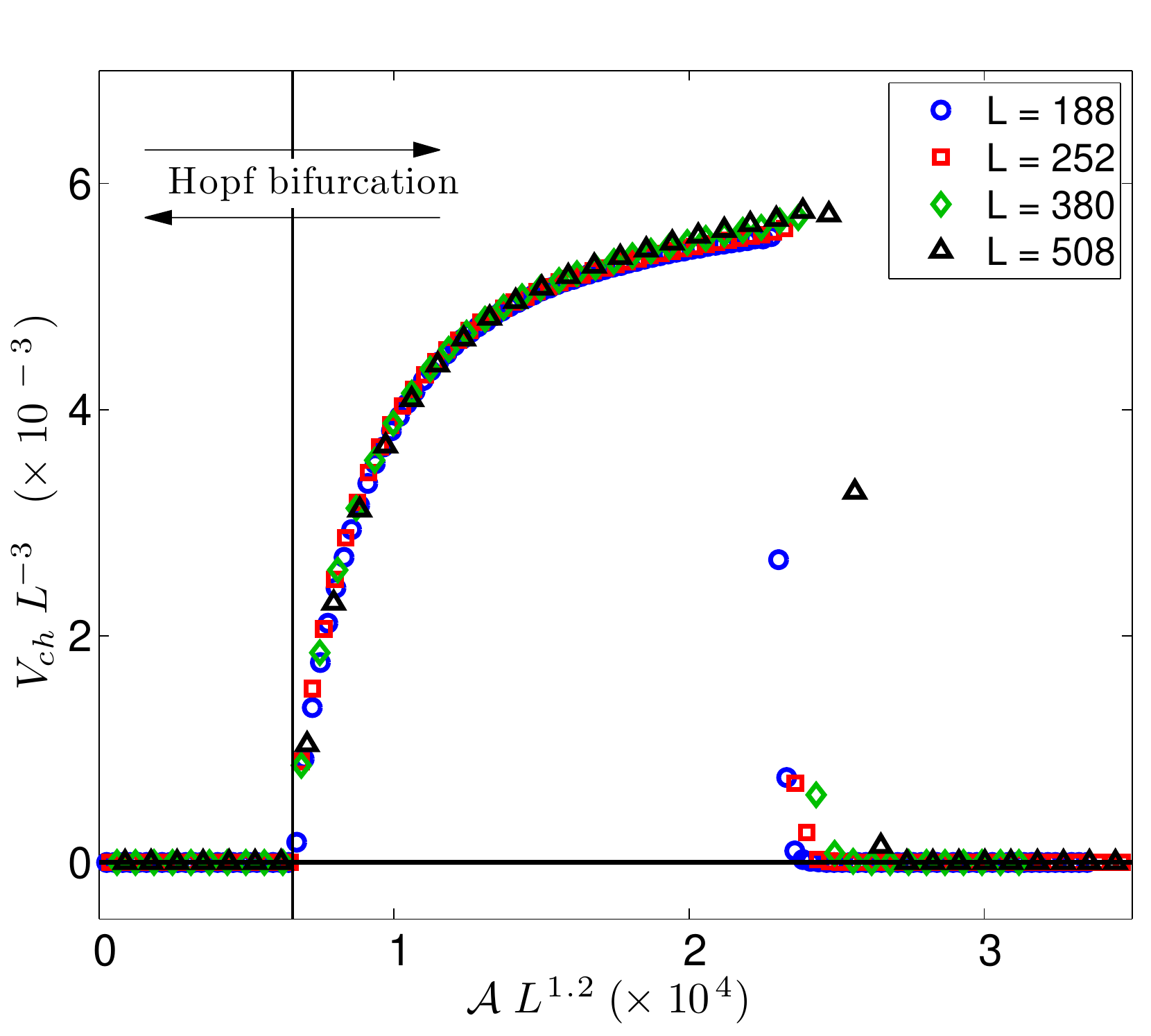} \label{fig:convex_hull_volume_L_scaling}} 
  \subfigure{\includegraphics[width=0.49\textwidth]{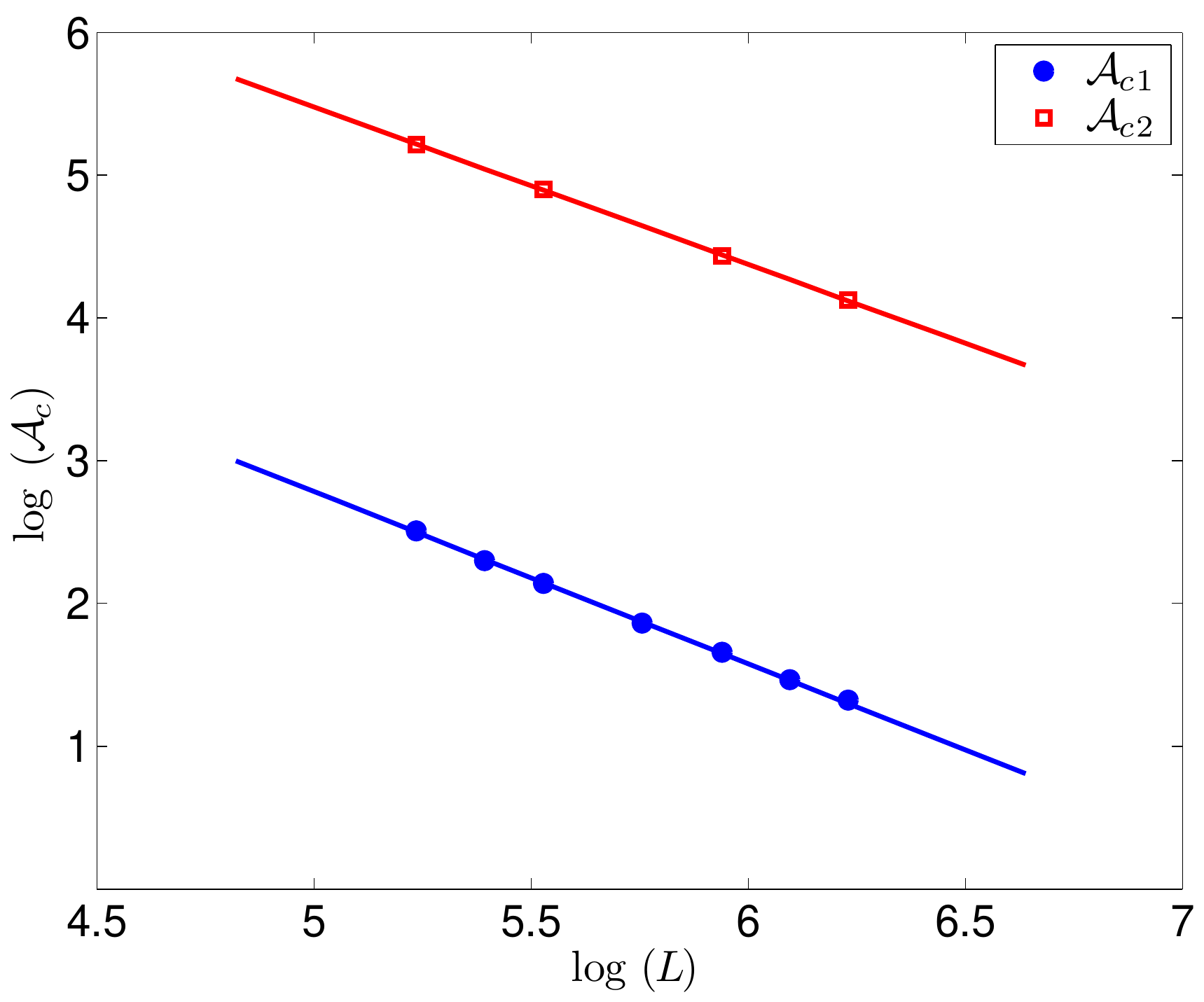} 
  \label{fig:Ac_L_scaling}}
  \caption{FIG.\ S3. (a) Variation of the rescaled volume of the convex hull $V_{\mrm{ch}}$ of the filament with $\mathcal{A}$ showing the transition between aplanar rotations in the regime $\mathcal{A} < \mathcal{A}_{c2}$ and planar oscillations in the regime $\mathcal{A} > \mathcal{A}_{c2}$. (b) Variation of the transition points $\mathcal{A}_{c1}$ and $\mathcal{A}_{c2}$ with $L$ exhibiting a scaling relation $\mathcal{A}_c \sim L^{\beta}$. Using Bayesian regression, we estimate $\beta = -1.2$ for $\mathcal{A}_{c1}$ and $\beta = -1.1$ for $\mathcal{A}_{c2}$, the different values responsible for the imperfect data collapse near the second transition. Symbols represent simulation data while solid lines represent the Bayesian estimate.
 } 
  \end{center}
\end{figure*}

\newpage
%
\end{document}